\documentclass [12pt]{jpsj3}
\usepackage{txfonts}
\usepackage[T1]{fontenc}
\usepackage{pifont}
\usepackage{subfigure}

\title{Multifractal Wave Functions of a System with a Monofractal Energy Spectrum}

\author{Masayuki Tashima\thanks{E-mail address : tashima@iis.u-tokyo.ac.jp} and Shuichi Tasaki
}
\inst{Department of Applied Physics, Waseda University, 3-4-1 Okubo, Shinjuku-ku, Tokyo 169-8555, Japan 
}
\abst{
We show the appearance of multifractal wave functions on a one-dimensional quasiperiodic system that has a monofractal energy spectrum.
Using the Mantica technique, we construct the model as an inverse problem from the energy spectrum of a \textit{pure} Cantor set. 
A relation between the critical state and the information dimension is proved and it is applied to the finite-size multifractal analysis.
}

\kword{quasiperiodic system, multifractal wave function, inverse problem}

\begin{document}
\maketitle

\section{INTRODUCTION}
Recently, the progress of experimental studies reveals interesting nature of the quasicrystal and new applications of the quasicrystal are expected to contribute to resolving a variety of problems \cite{macia06}. 
Under the circumstances of such experimental progress, theoretical study needs further development further to explain physical properties of the quasicrystal.
Some previous studies have revealed interesting nature of quasicrystals and quasiperiodic systems as a theoretical model of the quasicrystal. 
Strange nature of wave functions is an especially attractive problem, for example a critical state and a multifractal wave function as an eigenstate. \cite{fujiwara89} \cite{kohmoto87}.  

A fractal structure is also obtained in the energy spectrum of quasiperiodic systems. 
For example, the Fibonacci lattice is well known to have a Cantor-set-\textit{like} spectrum (with zero Lebesgue measure and multifractality \cite{kohmoto87}).
It raises another question as to whether the energy spectrum is always fractal in quasiperiodic Hamiltonians and vice versa. 
This relation is  an interesting problem, because the fractal dimensionality of the energy spectrum is connected to some physical properties of quasiperiodic systems, for instance, the temporal autocorrelation function \cite{ketzmerick92}. 
Some study showed the above relation in a particular model \cite{suto89}, but no general proof is available. 

One of the difficulties of theoretical study may be caused by the structure of quasiperiodic systems. 
A quasiperiodic function is designed to be a function that can be uniformly approximated by a Fourier series with a \textit{finite} number of pairwise incommensurate base frequencies \cite{baake02}.
Because of the incommensurability, quasiperiodic functions never have the translational periodicity and need a special construction technique.
Well known ways include the inflation rule \cite{kohmoto87}, an incommensurate potential, and the projection method \cite{quasicrystals08}. 
They were used in many previous works, for instance the Fibonacci lattice, the Thue-Morse lattice, the Harper model (1D), and the Penrose tiling (2D). 
They succeeded in explaining some aspects of the quasiperiodic system. 
However, the absence of simple symmetries of crystals makes further study of quasiperiodic systems difficult, and a new approach to studying quasiperiodic system is strongly needed.

Mantica proposed a completely different way, an inverse-problem approach. \cite{mantica96} \cite{mantica97} 
The Mantica technique is a way of constructing a quasiperiodic system from a multifractal spectrum. 
It offers us another way of studying quasiperiodic systems. 
Gurneri and Mantica used the technique and revealed the relation between the fractal energy spectrum and the anomalous diffusion of wave packets \cite{guarneri94}. 

In the present paper, we construct a one-dimensional quasiperiodic system with a \textit{mono}fractal energy spectrum by the Mantica technique and study the \textit{multi}fractality of its wave functions. 
The Mantica technique enables us to obtain an eigenstate of the singular continuous spectrum more easily than ordinary ways because we first set the energy spectrum and then calculate the corresponding Hamiltonian. 
We choose a pure Cantor set (with zero Lebesgue measure and monofractality) as the energy spectrum.
We conjecture that the dimension of the monofractal spectrum is related to the fractal dimension of the wave functions.
Moreover, we prove the relation between the information dimension and the index of the critical wave function.
Using the relation, we evaluate the multifractal spectrum for a finite number of lattice systems.

\section{MANTICA TECHNIQUE}
We construct a semi-infinite one-dimensional quasiperiodic system as an inverse problem through the Mantica technique.
It enables us to construct quasiperiodic systems from some kinds of multifractal spectra. 
Before presenting our results, we briefly review the Mantica technique introduced in the 1990s.  \cite{mantica96} \cite{mantica97} 
This is a technique of constructing a tridiagonal Hamiltonian possessing a given singular continuous spectrum belonging to the class of \textit{iterated function systems} (IFS). 

The Schr\"{o}dinger equation for a semi-infinite one-dimensional tight-binding model is generally given by 
\begin{equation}
 E\psi_n(E)=t_{n+1}\psi_{n+1}(E)+\epsilon_n \psi_n(E)+t_n \psi_{n-1}(E) , \label{eq:schro}
\end{equation} 
where $\psi_{n}(E)$ is the wave function at site $n$ with the energy $E$, $\epsilon_{n}$ is the on-site potential, and $t_{n}$ is the hopping element between the sites $n-1$ and $n$. 
This equation constitutes a recurrence formula, so that $\psi_{n}$ for energy $n$ can be calculated from deciding the first two terms and all coefficients $\{ \epsilon_{n} , t_{n} \}$. 

On the one hand, a set of orthogonal polynomials $\{  P_{n} (x) \}_{n}^{\infty} $ with respect to a measure $\mu (x)$ ($\mu (0) = 0$ and $\mu (1) = 1$), 
\begin{equation}
\int_{0}^{1} P_{n}(x) P_{m}(x) d \mu(x) = \delta_{nm} , \label{eq:orthogonal}
\end{equation}
satisfies the same form of equation \cite{polynomial}
\begin{equation} 
 xP_n(x)=b_{n+1}P_{n+1}(x)+a_n P_n(x)+b_nP_{n-1}(x) , \label{eq:3term}
\end{equation}
where 
\begin{equation*}
a_{n} = \int_{0}^{1} x P_{n}(x)^{2} d \mu (x)
\end{equation*}
and 
\begin{equation*}
b_{n} = \int_{0}^{1} x P_{n-1}(x) P_{n}(x) d \mu (x) .
\end{equation*}
We can find the following correspondence by comparing Eq. \eqref{eq:schro} to Eq. \eqref{eq:3term}:
\begin{gather*}
\psi_{n} \leftrightarrow P_{n} \\
E \leftrightarrow x \\ 
\mu(E) \leftrightarrow \mu (x) \\
\epsilon _{n} \leftrightarrow a_{n} \\
t_{n} \leftrightarrow b_{n} , 
\end{gather*}
where $\mu(E)$ is the integrated density of states.
Consequently, solving Eq. \eqref{eq:schro} is replaced by solving the three-term recurrence formula \eqref{eq:3term}. 

In general, computing a whole set of orthogonal polynomials $\{P_n\}$ associated with the measure $\mu$ is not an easy endeavor, but with the Mantica technique we can calculate the polynomials $\{P_n\}$ by using the invariability of IFS under an appropriate affine transformation
\begin{equation}
x \rightarrow \delta x + \beta \label{eq:affine} ,
\end{equation}
where $\delta$ and $\beta$ are real constants. The integral with a measure $\mu(x)$ satisfies
\begin{equation}
\int_{0}^{1} f (x) d \mu (x) = \sum_{i} \pi_{i} \int_{0}^{1} f (\delta_{i} x + \beta_{i}) d \mu (x) \label{eq:affine2} ,
\end{equation}
where the weights $\pi_{i}>0$ with $\sum_{i} \pi_{i} = 1$ for any continuous function $f$. 
The parameters $\delta_{i}$, $\beta_{i}$, and $\pi_{i}$ are determined from a preassigned spectral measure $ \mu (x)$ that belongs to IFS. 
In the Mantica technique, the parameters $a_{n} (= \epsilon_{n})$ and $b_{n} (= t_{n})$ can be calculated through Eq. \eqref{eq:affine2} and the polynomials $P_{n}$ are also calculated from the initialization $P_{-1} = 0$ and $P_{0} = 1$ \cite{mantica96}. 
We can thereby obtain the wave function $\psi_{n}(E)$ for an integrated density of states $\mu(E)$ that belongs to the class of IFS.

We calculate the first-order polynomial $P_{1}(x)$ as an example.
The first-order polynomial $P_{1}(x)$ must has the form
\begin{equation} 
P_{1}(x) = C_{1} x + C_{0} \label{eq:P1} ,
\end{equation}
where $C_{0}$ and $C_{1}$ are real constants.
The orthogonal relations for the first polynomial are given by 
\begin{equation}
\int_{0}^{1} P_{0}(x)P_{1}(x) d \mu (x) = C_{1} \langle x \rangle + C_{0} = 0 \label{eq:P0P1}
\end{equation}
and
\begin{equation}
\int_{0}^{1} P_{1}(x) P_{1} (x) d \mu (x) = {C_{1}}^{2} \langle x^{2} \rangle + 2 C_{1} C_{0} \langle x \rangle + {C_{0}}^{2} = 1 \label{eq:P1P1} .
\end{equation}
Here $\langle x^{n} \rangle$ is 
\begin{equation}
\langle x^{n} \rangle = \int_{0}^{1} x^{n} d \mu (x) \label{eq:x_n} .
\end{equation}
Thus the constants are given by 
\begin{equation}
C_{1} = \frac{1}{\sqrt{\langle x^{2} \rangle - {\langle x \rangle}^{2}}} \label{eq:C1}
\end{equation}
and 
\begin{equation}
C_{0} = C_{1} \langle x \rangle \label{eq:C0} .
\end{equation}
We can calculate the constants by determining $\langle x^{2} \rangle$ and $\langle x \rangle$.
They are obtained by the affine transformation Eq. \eqref{eq:affine2} as follows:
\begin{equation}
\langle x \rangle = \int_{0}^{1} x d \mu (x) = \sum_{i} \pi_{i} \int_{0}^{1} \left( \delta_{i} x + \beta_{i} d \mu (x) \right) = \sum_{i} \pi_{i} \left( \delta_{i} \langle x \rangle + \beta_{i} \right) .
\end{equation}
Therefore $\langle x \rangle$ is described by the parameters $\pi_{i}$, $\beta_{i}$, and $\delta_{i}$
\begin{equation}
\langle x \rangle = \frac{\sum_{j} \pi_{j} \beta_{j}}{1 - \sum_{i} \pi_{i} \delta_{i}} \label{eq:<x>} .
\end{equation}
We can determine $\langle x^{2} \rangle$ in the same way:
\begin{equation}
\langle x^{2} \rangle = \frac{\sum_{j} \left( {\beta_{j}}^{2} + 2 \delta_{j} \beta_{j} \langle x \rangle \right) }{1 - \sum_{i} \pi_{i} \delta_{i}} \label{eq:<x2>} .
\end{equation}
As a result, the first-order polynomial $P_{1}(x) = C_{1} x + C_{0}$ is obtained from the parameters $\pi_{i}$, $\beta_{i}$, and $\delta_{i}$. 
The higher-order polynomials $P_{n}(x)$ ($=\psi_{n}(E)$) are also computed in the same manner.

\begin{figure}
 \begin{center}
 \includegraphics[width=7cm,clip]{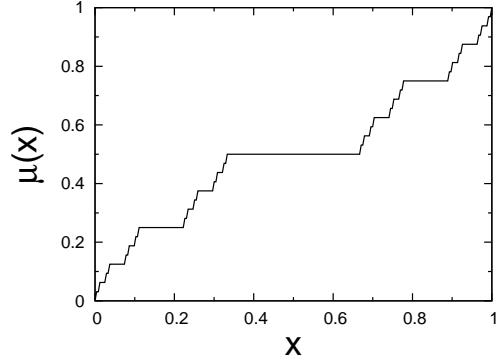}
 \end{center}
 \caption{The Devil's staircase.} 
 \label{fig:devil}
\end{figure}

In our study we choose as the energy spectrum the Devil's staircase (Fig. \ref{fig:devil}), which is described as IFS in the form,
\begin{equation}
  \mu(x) = \begin{cases}
  		 \frac{1}{2} \mu (3x) & (0 \le x < \frac{1}{3}) , \\
  		 \frac{1}{2} & (\frac{1}{3} \le x < \frac{2}{3}) , \\
  		 \frac{1}{2} \mu (3x-2) + \frac{1}{2} & (\frac{2}{3} \le x < 1) ,
  		   \end{cases} \label{eq:Cantor}
\end{equation}
where $x$ is the energy normalized in [0,1].
This corresponds to a \textit{pure} Cantor set (middle third removed).
This is a monofractal, since the generalized dimension $D_{q}$ of the measure $\mu$ is $D_{q} = D_{0} = \log 2 / \log 3  \simeq 0.631$. 
The parameters for the affine transformation \eqref{eq:affine2} are derived from the spectrum \eqref{eq:Cantor} as 
\begin{equation}
 \begin{split}
\int_{0}^{1} f(x) d \mu (x) &= \int_{0}^{1/3} f(x) d \mu (x) + \int_{1/3}^{2/3} f(x) d \mu (x)  \int_{2/3}^{1} f(x) d \mu (x) \\
                            &= \int_{0}^{1} f \left( \frac{x}{3} \right) d \mu \left( \frac{x}{3} \right) + \int_{0}^{1} f \left( \frac{x+1}{3} \right) d \mu \left( \frac{x+1}{3} \right) + \int_{0}^{1} f \left( \frac{x+2}{3} \right) d \mu \left( \frac{x+2}{3} \right) \\
                            &= \int_{0}^{1} \left( \frac{1}{2} f \left( \frac{x}{3} \right) + \frac{1}{2} f \left( \frac{x+2}{3} \right) \right) d \mu (x) \label{eq:parameters}.
 \end{split}
\end{equation}
Thus, the parameters are $\delta_{1} = \delta_{2} = 1/3 $, $\pi_{1}=\pi_{2}=1/2$,  $\beta_{1}=0$, and $\beta_{2}=2/3$.

\section{QUASIPERIODICITY OF THE TIGHT-BINDING MODEL}
In this section, we construct a one-dimensional system from the Devil's staircase and confirm its quasiperiodicity.
The semi-infinite lattice system can be obtained from the recursive sequence of the Mantica technique, but calculating a very large number of polynomials corresponding to the number of lattice sites is a hard work for computer.
Hence we choose $2^{13}=8192$ as the site number. 
Setting the site number finite never changes  the boundary condition \cite{2^13}.  

\begin{figure}
 \begin{center}
 {\includegraphics[width=7cm,clip]{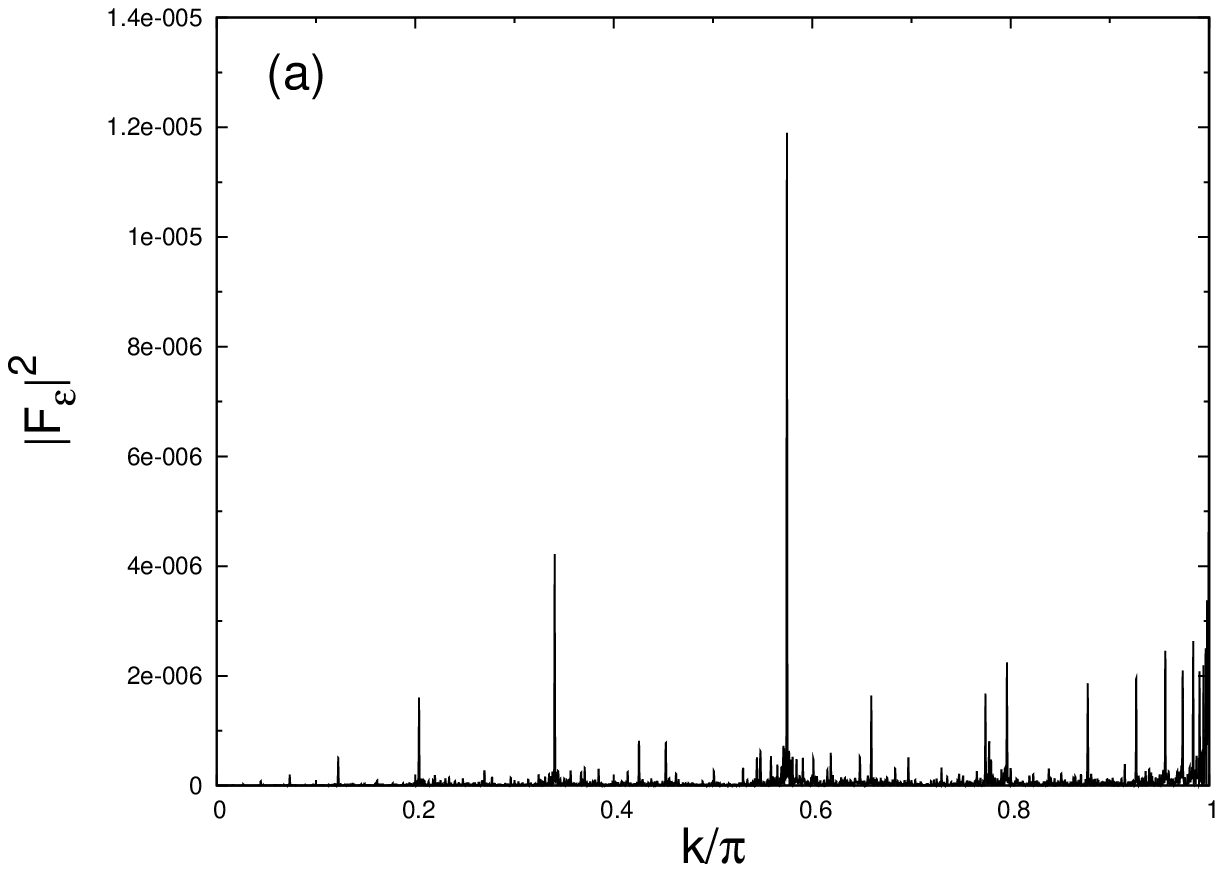}\label{fig:Pspct(cantor)}}
 \hspace{5mm}
 {\includegraphics[width=7cm,clip]{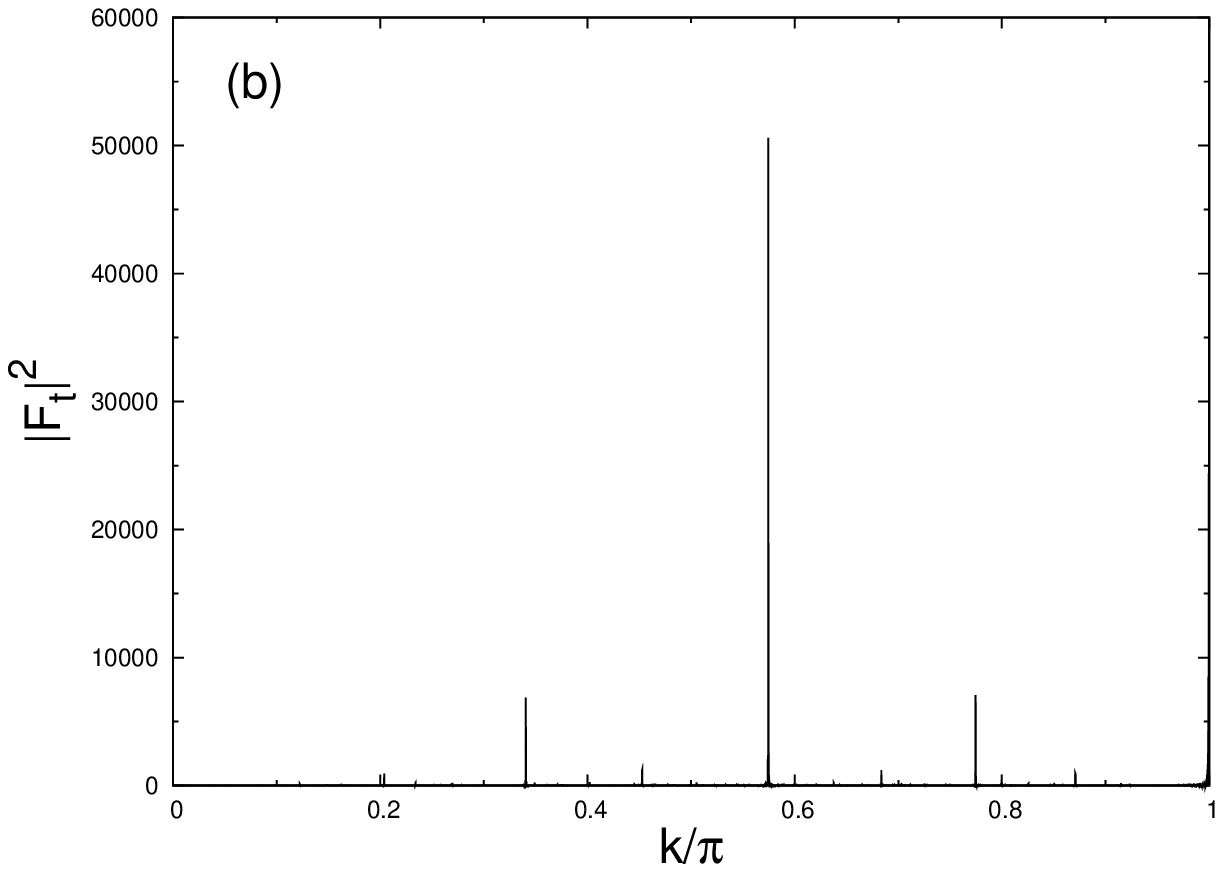}\label{fig:Hspct(cantor)}}  
 \end{center}
 \caption{The power spectra of the (a) on-site potential $\epsilon_{n}$ and (b) the hopping element $t_{n}$. The power spectrum of hopping element $t_n$. Since peaks are at $k/\pi = 0.5741 \dots , 0.7745 \dots , 0.3404 \dots $, the Fourier spectrum is not described by a simple proportional relation. A clear hierarchical structure is also observed.} 
 \end{figure}

First we investigate the power spectrum of the on-site potential $\epsilon_n$ and the hopping element $t_n$ in order to examine the quasiperiodicity of the system. 
First 50 data points are excluded so as to eliminate the boundary effect at $n=0$. 
The Fourier spectrum of $\epsilon_n$ and $t_n$ are given by
\begin{equation}
 |F_{\epsilon}|^2 = \Bigl| \sum_n (\epsilon_n - \langle \epsilon \rangle ) e^{i n k} \Bigr| ^2  \label{eq:e-fourier}
\end{equation}
and
\begin{equation}
 |F_{t}|^2 = \Bigl| \sum_n (t_n - \langle t \rangle ) e^{i n k} \Bigr| ^2  \label{eq:t-fourier},
\end{equation}
where $\langle \epsilon \rangle$ and $\langle t \rangle$ are the arithmetic means of $\epsilon_n$ and $t_n$, respectively (see Fig. \ref{fig:Pspct(cantor)} and Fig. \ref{fig:Hspct(cantor)}).
The spiky structure indicates a long range order.
The positions of the peaks without a simple proportional relation, on the other hand, indicates that it has no periodic arrangement. 
We thereby claim that the system is quasiperiodic \cite{caution}. 

The potential and the hopping element of our system do not take only two values in real space, while the Fibonacci lattice  and the Thue-Morse lattice consist of two elements as the potential or the hopping (the quasiperiodicity only appears in its arrangement). 
The similarity between our system and these quasiperiodic systems is unfortunately not observed from the viewpoint of Hamiltonian elements.   
Moreover, the potential of our model is not described by a cosine curve with an irrational number, such as in the Harper model.


\section{MULTIFRACTAL ANALYSIS OF CRITICAL STATE}
In present section, we describe how we can characterize the fractality of the wave functions of our system.
Wave functions are generally classified into three types, namely \textit{extended}, \textit{localized}, and \textit{critical} \cite{ppq}.
An extended state is defined as
\begin{equation}
\int_{|r|<L} |\psi (\vec{r})|^{2} d \vec{r} \sim L^{D} \label{eq:e_state} ,
\end{equation}
where $L$ is the system size and $D$ is the spatial dimension. 
It is similar to the Bloch state.
A localized state is 
\begin{equation}
\int_{|r|<L} |\psi (\vec{r})|^{2} d \vec{r} \sim L^{0} \label{eq:l_state} .
\end{equation}
The last type of the wave function, the critical state is neither localized nor extended:
\begin{equation}
\int_{|r|<L} |\psi (\vec{r})|^{2} d \vec{r} \sim L^{\nu} \label{eq:c_state} ,
\end{equation}
where the index $\nu$ is in the range $0 < \nu < D$.
A typical wave function in the critical state may be a power-law type function $\psi(\vec{r}) \sim |\vec{r}|^{\nu}$ with $\nu < D / 2$ or a self-similar function \cite{ppq}. 

Equation \eqref{eq:c_state} can be rewritten for our one-dimensional discrete system as 
\begin{equation}
S(N) \equiv \sum_{i=0}^{N} |\psi_{i}|^{2} \sim N^{\nu_{N}} \label{eq:discrete_c} ,
\end{equation}
where $N$ is the site number and $\nu_{N}$ is a value of the index $\nu$ for the $N$-site system. 
The logarithm of Eq. \eqref{eq:discrete_c} leads
\begin{equation}
\ln S(N) \sim \nu_{N} \ln N \label{eq:ln_S} .
\end{equation}
We can estimate the index $\nu_{N}$ in the range $(0,1)$ as a gradient in a plot of $\ln S(N)$ against $\ln N$. 

Another way of characterizing the critical state is the multifractal analysis. 
The multifractality is represented by the singularity index, or the Lipschitz-H\"{o}lder exponent, $\alpha_{q}$, and the multifractal spectrum $f(\alpha_{q})$. \cite{fractals} \cite{feder}
The singularity index $\alpha_{q}$ describes  the local degree of singularity (local fractal dimension) and the multifractal spectrum $f(\alpha_{q})$ is the fractal dimension of the support which has the singularity $\alpha_{q}$.
In other words, the curve $f(\alpha_{q})$ against $\alpha_{q}$ means a relation between the local fractal dimension characterizing the system and the global fractal dimension of the spatial distribution of the singularity. 

The multifractal analysis has been applied to study of the nature of wave functions.
The analysis shows that an extended state has $f(\alpha_{q} = 1) = 1$, a localized state has $f(\alpha_{q} = 0) = 0$ and $f(\alpha_{q} = \infty) = 1$, and a critical state only has a smooth convex multifractal spectrum $f(\alpha_{q})$ in the range $[\alpha_{\rm{min}}, \alpha_{\rm{max}}]$.
For example, the multifractal spectrum for a Fibonacci lattice was obtained by T. Fujiwara \textit{et al} \cite{fujiwara89}.
They obtained an asymmetric shape of $f(\alpha_{q})$ and the limits $f(\alpha_{\rm{max}} = \alpha_{- \infty}) = 0$ and $f(\alpha_{\rm{min}} = \alpha_{+ \infty}) \neq 0$ for the wave functions at the edge of the energy spectrum. 
The multifractal spectra $f(\alpha_{q})$ of the critical states of other quasiperiodic systems are also smooth.

In the present paper, we consider a discrete lattice system, and therefore the multifractal formalism is modified to the following equations.
The spatial distribution $Q_{i}$ (which is described as the probability measure in mathematics) at cell $i$ with size $l$ is given by 
\begin{equation}
Q_{i} \left( \frac{l}{N} \right) = \frac{ \sum_{k=i \, l}^{(i + 1) \, l -1} |P_{k} (x)|^{2} }{ \sum_{j=1}^{N} |P_{j} (x)|^{2} } \label{eq:Q_i} ,
\end{equation}
where the summation in the denominator is for normalization and $N$ is the number of the sites on the whole lattice. 
The participation function $Z_{q}$ is defined by
\begin{equation}
Z_{q} \left( \frac{l}{N} \right) = \sum_{i=1}^{m} \left[ Q_{i} \left( \frac{l}{N} \right) \right]^{q} \label{eq:Z_q} ,
\end{equation}
where $m = N / l$ is the number of the cells.
In Eq. \eqref{eq:Z_q}, different parts of the distribution $Q_{i}$ can be stressed by the parameter $q$. 
If $q$ has a large positive value, the participation function $Z_{q}$ is dominated by the regions corresponding to larger values of $Q_{i}$, whereas if $q$ is a negative large number, $Z_{q}$ is dominated by contributions from small values of $Q_{i}$.
The generalized dimension $D_{q}$ is given by
\begin{equation}
D_{q} = \frac{1}{q-1} \lim_{N \to \infty} \lim_{\epsilon \to \epsilon_{\rm{min}}} \frac{\ln Z_{q}(l/N)}{\ln \epsilon} \label{eq:D_q} .
\end{equation}
A lattice system has the lattice constant as the minimal value, and therefore the extremum $\epsilon$ has to be fixed at the lower cut off length scale $\epsilon_{\rm{min}}=l/N=2/N$ for our system.
The singularity index $\alpha_{q}$ and the multifractal spectrum $f(\alpha_{q})$ are calculated from the $D_{q}$ as 
\begin{equation}
\alpha_{q} = \frac{d}{dq} \Bigl[ \left( q-1 \right) D_{q} \Bigr] \label{eq:alpha_q} 
\end{equation}
and
\begin{equation}
f(\alpha_{q}) = q \alpha_{q} - (q-1) D_{q} \label{eq:f_alpha} .
\end{equation}
As is easily seen from Eq. \eqref{eq:alpha_q} and Eq. \eqref{eq:f_alpha}, the multifractal spectrum for a monofractal reduces to the dimension of the monofractal $D_{f} = f(\alpha_{q}) = \alpha_{q}$.

If $q=1$, the generalized dimension $D_{q}$ is called the information dimension $D_{1}$
\begin{equation}
D_{1} = \lim_{N \to \infty} \frac{1}{\ln (l / N)} \sum_{i=1}^{m} Q_{i} \left( \frac{l}{N} \right) \ln Q_{i} \left( \frac{l}{N} \right).
\end{equation}
The information dimension $D_{1}$ coincides with the index of the critical state $\nu_{N}$ given in \eqref{eq:discrete_c} (see Appendix).
The information dimension for the finite number of lattice system $D_{1}^{N}$ is described as
\begin{equation}
D_{1}^{N} = \nu_{N} + \delta D_{1} (N) , \text{where} \lim_{N \to \infty} \delta D_{1}(N) = 0 \label{eq:delta_D1}.
\end{equation}
We can evaluate the multifractal spectrum for the finite number of lattice system through the $n$ dependence of $D_{1}^{N} - \nu_{N}$.

\section{RESULTS AND DISCUSSION}
\begin{figure}
 \begin{center}
 \includegraphics[width=7cm,clip]{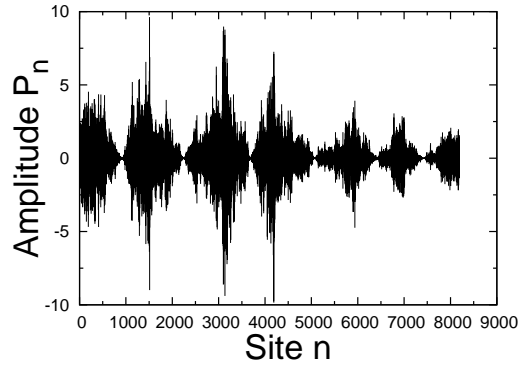} 
 \end{center}
 \caption{Eigenfunction for the energy $x = 1/3$. There are five clusters except for the clusters around the boundary sites $n=0$ and $n=2^{13} = 8192$. }
 \label{fig:wf(cantor_x..332)}
\end{figure}

A typical wave function is illustrated in Fig. \ref{fig:wf(cantor_x..332)}.
It clearly lacks periodicity.
To characterize the structure of the wave function, we study the critical state and the multifractality.
First 50 data points are also excluded from both analysis.

\begin{figure}
 \begin{center}
 \includegraphics[width=7cm,clip]{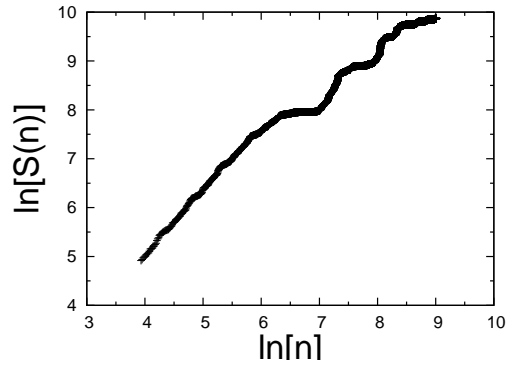} 
 \end{center}
 \caption{$S(N)$ for the eigenfunction for the energy $x = 1/3$.}
 \label{fig:sn_wf(cantor_x..332)}
\end{figure}

Let us first describe the analysis given in Eq. \eqref{eq:ln_S}.
We evaluate $S(N)$ for the wave function illustrated in Fig. \ref{fig:wf(cantor_x..332)}.
The resulting plot in Fig. \ref{fig:sn_wf(cantor_x..332)} yields the index $\nu_{N} \simeq 0.886$. 
It indicates that  this wave function is in the critical state.

Wave functions in the critical state may be a power-law type function or a self-similar function \cite{ppq}.
The wave function illustrated in Fig. \ref{fig:wf(cantor_x..332)} might be a series of power-law-type localized functions.
In order to check the possibility, we divided the wave function into five \textit{clusters}, where a cluster is defined as a region of large amplitudes between sites of small amplitudes as indicated in Fig. \ref{fig:wf(cantor_x..332)} and applied the power-law approximation to each cluster.
However, the clusters were not well described by the approximation.
In addition, the function in Fig. \ref{fig:wf(cantor_x..332)} is not a simple self-similar function, that was reported on a Fibonacci lattice \cite{fujiwara89}.
Hence, we analyzed them from the other viewpoint, namely the multifractal analysis.

\begin{figure}
 \begin{center}
 \includegraphics[width=8cm,clip]{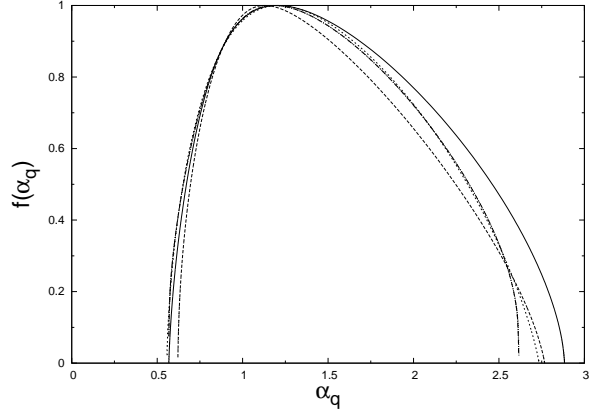} 
 \end{center}
 \caption{The multifractal spectra $f(\alpha_{q})$ are plotted for the wave functions with the energies $x=1/27$ (solid line), $x=1/9$ (broken line), $x=2/9$ (long dashed line), $x=1/3$ (dotted line), and $x=2/3$ (dash-dotted line).}
 \label{fig:f_alpha_cantor}
\end{figure}

We then investigate the multifractality of the wave functions. 
Figure \ref{fig:f_alpha_cantor} shows the multifractal spectrum of the wave functions of the system.
We thereby conclude that the wave function of the system with a monofractal energy spectrum exhibits multifractality.
The maximum value of $f(\alpha_{q})$, the dimension of the support, is unity.
The equivalence of the spatial dimension and the dimension of the support means that there is no areas with zero amplitude. 
We calculate the average of the limitation values of the multifractal spectrum as $f(\alpha_{\rm{min}}) \simeq 0.038$ at $\alpha_{\rm{min}} \simeq 0.58$ and $f(\alpha_{\rm{max}}) \simeq 0$ at $\alpha_{\rm{max}} \simeq 2.7$.
The multifractal spectrum $f(\alpha_{q})$ exhibits the maximum value unity at $\alpha_{q} \simeq 1.2$. 
Any simple relations between the dimension of the energy spectrum $D_{\rm{cantor}} = \log 2 / \log 3 \simeq 0.631$ and the multifractal spectrum $f(\alpha_{q})$ are not observed in Fig. \ref{fig:f_alpha_cantor}.

The multifractal analysis for the finite-size lattice system is needed to be treated with discretion \cite{fujiwara89}.
Thus we confirmed the convergence of the multifractal spectrum $f(\alpha_{q})$ by Eq. \eqref{eq:delta_D1}.
\begin{figure}
 \begin{center}
 \includegraphics[width=8cm,clip]{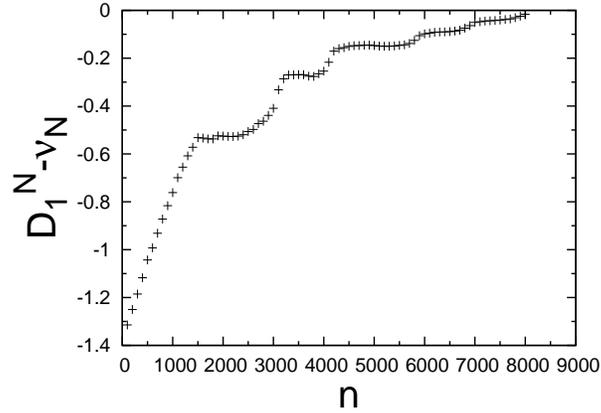} 
 \end{center}
 \caption{The declination $D_{1}^{N} - \nu_{N}$ against the number of site $n$ for the energy $x=1/3$.}
 \label{fig:D1-nu}
\end{figure} 
We can see the convergence of $D_{1}^{N} - \nu_{N}$ against $N$ in Fig. \ref{fig:D1-nu}.

\section{SUMMARY}
We constructed a quasiperiodic system from the pure Cantor set spectrum by the Mantica technique and proved the relation between the critical state and the information dimension. 
The preassigned spectrum is monofractal.
Nevertheless, the wave functions of our system exhibits the multifractality. 
In this work, we have not observed a simple relation between the dimension of the energy spectrum and the multifractal spectrum of the wave functions.
More study is needed to understand the fractal nature of wave functions of quasiperiodic systems. 

\section*{Acknowledgements}
We are indebted to M. Fujiyoshi and S. Ajisaka for very useful discussions. 
It is a pleasure to thank Y. Ishii for fruitful comments and information about this work.
We wish to thank N. Hatano for recommendation to publish this paper.
This work is supported by Grant-Aid for Scientific Research No. 21540398 from the Ministry of Education, Culture, Sports, Science and Technology as well as by Core Research for Evolutional Science and Technology (CREST) of Japan Science and Technology Agency.

Lastly, one of the authors, Shuichi Tasaki passed away before publishing this paper. 
The other author, M.T. really appreciates his great guidance and dedicates this work to his memory.

\appendix
\section{A RELATION BETWEEN THE CRITICAL STATE AND THE INFORMATION DIMENSION}
The critical state for a finite number of lattice sites $N$ is defined as 
\begin{equation}
\sum_{i=1}^{N} {\lvert \psi_{i} \rvert}^{2}  = C_{N} \, N^{\nu_{N}} \label{eq:Nall},
\end{equation}
where $\psi_{i}$ is the wave function at site $i$, $C_{N}$ is a positive constant and, $\nu_{N}$ is the index for $N$-sites lattice system.
We suppose that $\nu_{N}$ coincides with $\nu$ in the limiting case
\begin{equation}
\lim_{N \to \infty} \nu_{N} = \nu .
\end{equation} 
We divide the whole system into $m$ parts with size $l=N/m$ and assume that the summation of the wave functions of $i$th cell $N_{i}$ satisfies 
\begin{equation}
\sum_{j}^{N_{i}} {\lvert \psi_{j} \rvert}^{2}  = C_{i} \, l^{\nu_{i}} \label{eq:Ni},
\end{equation}
where $C_{i}$ is a positive constant and $\nu_{i}$ is a critical index of the $i$th cell.
The summation of Eq. \eqref{eq:Ni} over $m$ parts should reduce to Eq. \eqref{eq:Nall}:
\begin{equation}
\sum_{i}^{m} \sum_{j=1}^{N_{i}} {\lvert \psi_{j} \rvert}^{2} = \sum_{i=1}^{m} C_{i} \, l^{\nu_{i}} = C_{N} \, N^{\nu_{N}} .
\end{equation}

The spatial distribution $Q_{i}$ at the $i$th cell is given by
\begin{equation}
Q_{i} \left( \frac{l}{N} \right) = \frac{\sum_{N_{i}} {\lvert \psi \rvert}^{2}}{\sum_{N} {\lvert \psi \rvert}^{2}} = \frac{C_{i} \, l^{\nu_{i}}}{C_{N} \, N^{\nu_{N}}} 
\end{equation}
and $Q_{i}$ is normalized as 
\begin{equation}
\sum_{i=1}^{m} Q_{i} \left( \frac{l}{N} \right) = \frac{1}{C_{N} \, N^{\nu_{N}}} \sum_{i=1}^{m} C_{i} \, l^{\nu_{i}}= 1 .
\end{equation}
The information dimension $D_{1}$ for the infinite lattice system is defined as
\begin{equation}
D_{1} = \lim_{N \to \infty} \frac{1}{\ln (l / N)} \sum_{i=1}^{m} Q_{i} \left( \frac{l}{N} \right) \ln Q_{i} \left( \frac{l}{N} \right).
\end{equation} 
The summation in the above equation is rewritten by
\begin{equation}
 \begin{split}
\sum_{i=1}^{m} Q_{i} \left( \frac{l}{N} \right) \ln Q_{i} \left( \frac{l}{N} \right) &= \sum_{i=1}^{m} \left( \frac{1}{C_{N} \, N^{\nu_{N}}} C_{i} \, l^{\nu_{i}} \right) \ln \left( \frac{1}{C_{N} \, N^{\nu_{N}}} C_{i} \, l^{\nu_{i}} \right) \\
                                                                                             &= \frac{1}{C_{N} \, N^{\nu_{N}}} \sum_{i=1}^{m} C_{i} \, l^{\nu_{i}} \ln \left( C_{i} \, l^{\nu_{i}} \right) - \frac{1}{C_{N} \, N^{\nu_{N}}}  \ln \left( C_{N} \, N^{\nu_{N}} \right) \sum_{i=1}^{m} C_{i} \, l^{\nu_{i}} . 
 \end{split} \label{eq:Q_lnQ} 
\end{equation}
The second term of Eq. \eqref{eq:Q_lnQ} reduces to 
\begin{equation}
\frac{1}{C_{N} \, N^{\nu_{N}}}  \ln \left( C_{N} \, N^{\nu_{N}} \right) \sum_{i=1}^{m} C_{i} \, l^{\nu_{i}} = \ln \left( C_{N} \, N^{\nu_{N}} \right) = \ln C_{N} + \nu_{N} \ln N .
\end{equation}
Therefore the information dimension $D_{1}$ is given by
\begin{equation}
 \begin{split}
D_{1} &= \lim_{N \to \infty} \frac{1}{\ln N - \ln l } \left[ \ln C_{N} + \nu_{N} \ln N - \frac{1}{C_{N} \, N^{\nu_{N}}} \left( \sum_{i=1}^{m} C_{i} \, l^{\nu_{i}} \ln C_{i} + \ln l \sum_{i=1}^{m} \nu_{i} C_{i} \, l^{\nu_{i}}  \right) \right] \\
           &= \nu + \lim_{N \to \infty} \frac{\ln C_{N} + \nu_{N} \ln l }{\ln N - \ln l } - \frac{1}{\ln N - \ln l } \frac{1}{C_{N} \, N^{\nu_{N}}
} \left( \sum_{i=1}^{m} C_{i} \, l^{\nu_{i}} \ln C_{i} + \ln l \sum_{i=1}^{m} \nu_{i} C_{i} \, l^{\nu_{i}}  \right) .
 \end{split} \label{eq:D1} 
\end{equation}

We now define ${ \left[ X_{i} \right] }_{\rm{max}}$ as the maximum value of $\{ X_{i} \}$ in order to evaluate a behavior of the absolute value $\lvert \ln C_{i} \rvert$:
\begin{equation}
\max_{i} \{ \lvert \ln {C_{i}} \rvert \} = { \left[ \lvert \ln {C_{i}} \rvert \right] }_{\rm{max}} .
\end{equation}
We then obtain the following inequality: 
\begin{equation}
\left \lvert \sum_{i=1}^{m} C_{i} \, l^{\nu_{i}} \ln C_{i} \right \rvert < \sum_{i=1}^{m} C_{i} \, l^{\nu_{i}} { \left[ \lvert \ln C_{i} \rvert \right] }_{\rm{max}} = { \left[ \lvert \ln C_{i} \rvert \right] }_{\rm{max}} \, C_{N} \, N^{\nu_{N}}
\end{equation}
Then then the third term of Eq. \eqref{eq:D1} converges to zero:
\begin{equation}
 \begin{split}
\lim_{N \to \infty} \frac{1}{\ln N - \ln l } \frac{1}{C_{N} \, N^{\nu_{N}}} \left \lvert \sum_{i=1}^{m} C_{i} \, l^{\nu_{i}} \ln C_{i} \right \rvert &< \lim_{N \to \infty} \frac{1}{\ln N - \ln l } \frac{1}{C_{N} \, N^{\nu_{N}}} { \left[ \lvert \ln C_{i} \rvert \right] }_{\rm{max}} C_{N} \, N^{\nu_{N}} \\
                            &= \lim_{N \to \infty} \frac{{ \left[ \lvert \ln C_{i} \rvert \right] }_{\rm{max}}}{\ln \left[ N \right] - \ln \left[  l \right]} = 0 .
 \end{split}
\end{equation}
The forth term in Eq. \eqref{eq:D1} is evaluated in the same manner.
It also converges to zero:
\begin{equation}
\lim_{N \to \infty} \frac{\ln l }{\ln N - \ln l } \frac{1}{C_{N} \, N^{\nu_{N}}} \sum_{i=1}^{m} \nu_{i} C_{i} \, l^{\nu_{i}} \le \lim_{N \to \infty} \frac{{\left[ \nu_{i} \right]}_{\rm{max}} \ln l }{\ln N - \ln l } = 0 .
\end{equation}
We thus arrive at the relation between the information dimension $D_{1}$ and the index $\nu$:
\begin{equation}
D_{1} (=\alpha_{1} = f(\alpha_{1})) = \nu \label{eq:D1=nu}.
\end{equation}
If a state is localized, $D_{1} = \nu = 0$ and if a state is extended, $D_{1} = \nu = 1$.
These results are consistent with a previous research \cite{ppq}.
We can obtain the same relation for a continuous system in the same way.

Finally, the information dimension for a finite lattice system, $D_{1}^{N}$, is given by
\begin{equation}
 \begin{split}
D_{1}^{N} = \nu_{N} &+ \delta D_{1}(N)\\
                   = \nu_{N} &+ \frac{1}{\ln N - \ln l } \frac{1}{C_{N} \, N^{\nu_{N}}} \left( \sum_{i=1}^{m} C_{N} \ln C_{N} \frac{N^{\nu_{N}}}{n} - C_{i} l^{\nu_{i}} \ln C_{i} \right) \\
                             &+ \frac{\ln l }{\ln N - \ln l } \frac{1}{C_{N} \, N^{\nu}} \left( \sum_{i=1}^{m} \nu_{N} \, C_{N} \frac{N^{\nu_{N}}}{n} - \nu_{i} \, C_{i} l^{\nu_{i}} \right) .
 \end{split}                           
\end{equation}
The each term in the brackets can be interpreted as representing a roughness of a local structure.






\begin{thebibliography}{99}
\bibitem{2^13} Site number $2^{13}$ is selected by G. Mantica in Ref. \citen{mantica97}. 
It does not have any specific meanings in mathematics and in physics. 
The limitation never changes any conditions, because the orthogonal polynomials $\{P_{n}\}$ are calculated by Eq. \eqref{eq:3term} from $n=1$ to $n=m$ $(m > 1)$. 
The polynomials $\{P_{n}\}$ for $n > 2^{13}$ can be also derived with more memory and time requirements.
\bibitem{caution} We used the discrete Fourier transformation. Therefore we cannot say for sure the number of peaks are really finite. However, numerical evidence seems to suggest quasiperiodicity.
\end{thebibliography}
\end{document}